\documentclass[a4paper,fleqn]{cas-sc}

\usepackage[authoryear,longnamesfirst]{natbib}
\usepackage{amsmath}
\usepackage{gensymb}
\usepackage{lineno}
\newcommand{\bfm}[1]{\mbox{\boldmath{$#1$}}}

\begin{document}
\let\WriteBookmarks\relax
\def\floatpagepagefraction{1}
\def\textpagefraction{.001}

\shorttitle{Unfavorable conditions in 1998 KY26}

\shortauthors{K. Hansen and M. Hirabayashi}

\title [mode = title]{Thermally unfavorable conditions of volatile existence in near-Earth asteroid 1998 KY26, the Hayabusa extended mission's target}

\author[1]{Korbin Hansen}[type=editor, orcid=0009-0004-2905-0957]
\cormark[1]
\ead{khansen49@gatech.edu}
\credit{Methodology, Software, Investigation, Data curation, Writing - original draft, Writing - review \& editing, Visualization}

\author[1]{Masatoshi Hirabayashi}[type=editor, orcid=0000-0002-1821-5689]
\ead{thirabayashi@gatech.edu}
\credit{Conceptualization, Methodology, Writing - review \& editing, Supervision, Project administration}

\affiliation[1]{organization={Daniel Guggenheim School of Aerospace Engineering, Georgia Institute of Technology}, addressline={270 Ferst Dr}, city={Atlanta}, postcode={30332}, state={GA},country={USA}}

\cortext[cor1]{Corresponding author}


\begin{abstract}
The near-Earth asteroid 1998 KY26, the target of the Hayabusa2\# extended mission, has been proposed as a dark comet that may host volatiles, particularly water, owing to an anomalous out-of-plane nongravitational acceleration that may result from outgassing. This asteroid, however, may also belong to a largely anhydrous taxonomic group. Whether outgassing-driven \textcolor{black}{acceleration is possible} strongly depends on how the asteroid stores volatiles in its interior and releases them upon heating at the surface. Thus far, the contentious topic of the existence of volatiles on 1998 KY26 that can cause outgassing has not been extensively examined from a \textcolor{black}{thermophysical standpoint}. Here, we show that all locations on the surface or in the interior of 1998 KY26 have orbit-averaged temperatures significantly above the sublimation point of volatiles in vacuum. Across 13 thermal characterizations, this observation is found to be consistent despite parameter variations in equivalent diameter, thermal conductivity, Bond albedo, specific heat capacity, and spin pole ecliptic latitude and longitude. \textcolor{black}{Our temperature model results indicate that the water outgassing rate on 1998 KY26 would be too high to maintain buried ice}, refining the understanding of the asteroid’s composition.
\end{abstract}

\begin{highlights}
\item Orbital-averaged temperature of 1998 KY26 far above sublimation threshold of ice
\item Calculated water loss rates preclude 1998 KY26 being a dark comet at present
\item Thermophysical parameter changes weakly affect average temperature of 1998 KY26
\end{highlights}


\maketitle

\section{Introduction}

The near-Earth asteroid 1998 KY26 is the target of the Japan Aerospace Exploration Agency's Hayabusa2\# extended mission with a planned rendezvous in 2031 \citep{hirabayashi2021hayabusa2, hirabayashi2026overview}. This asteroid is notable for its extremely quick rotation period. \cite{ostro1999radar} initially reported a rotation period of 10.70 minutes. Having resolved the spin period, \textcolor{black}{Doppler broadening measurements were used to bound the asteroid's diameter from 20 m to 40 m. Currently, 1998 KY26's rotation period is reported to be an even faster 5.35 minutes, as \cite{santana2025hayabusa2}} integrated observations over a larger range of phase angles to resolve complex patterns, resulting in the rotation period being halved. Using the relationship between radar bandwidth, object size, and spin period, this reduction in spin period evidences a commensurate reduction in equivalent diameter to 11 m. \textcolor{black}{Beyond its small size, 1998 KY26 is also notable for its high albedo. \cite{tatsumi2026hayabusa2} reports a geometric albedo of 0.55, which, alongside the asteroid's spectral slope, is consistent with an E-type (enstatite achondrite) taxonomy. The E-type class is thought to have igneous origins \citep{keil2010enstatite} and includes notable asteroids such as Nysa and Hungaria \citep{clark2004type}, along with other consistently reflective members.}

Among the ongoing debates regarding 1998 KY26 is whether it is a dark comet, a small object that dynamically behaves like a comet, exhibiting non-negligible non-gravitational forces without observations of mass ejection (dust and sublimation). Given a parametric model for these accelerations in terms of their radial ($A_1$), transverse ($A_2$), and normal ($A_3$) components \citep{marsden1973comets}, direct solar radiation pressure and the Yarkovsky effect commonly account for the $A_1$ and $A_2$ components, though recent work reported that 1998 KY26 has an unusually high $A_3$ term \citep{seligman2023dark}. One explanation is that this asteroid contains volatiles in its interior and releases them to produce the out-of-plane acceleration \citep{seligman2023dark}. 1998 KY26 characteristically lacks a coma, but the measured $A_3$ acceleration infers that, if volatile (water) sublimation is a major source, the production (outgassing) rate is approximately $3\times10^{-5}$ kg/s \citep{seligman2024two}. 

\textcolor{black}{Arguments against 1998 KY26 being a dark comet have come from multiple angles.} From a dynamical perspective, \textcolor{black}{an astrometric study found} that the measured high $A_3$ term could result \textcolor{black}{from the combination of direct solar radiation pressure and thermal emission (including the Yarkovsky effect)}, suggesting no need for the dark comet concept in this instance \citep{farnocchia2025radiation}. \textcolor{black}{Additionally, the asteroid's orbit, with an apoapsis of 1.475 AU and a periapsis of 0.983 AU \citep{SBDB1998KY26}, takes it close to the Sun.} At such heliocentric distances, the asteroid receives a higher heat flux and experiences higher temperatures. Under this orbital condition, volatiles may not remain in their solid phase, though burial may slow sublimation rates \citep{schorghofer2008lifetime}. From a geologic perspective, the violent formation conditions of the E-type asteroid class are thought to have rendered its constituents relatively anhydrous \citep{zhang2023compositional, fornasier2011spectroscopic}. Recent in-situ analyses of aubrites (E-type meteorite analogs) revised down the poor water content by 2 to 3 orders of magnitude \citep{peterson2023h}. \textcolor{black}{Furthermore, as 1998 KY26 is a fast rotator regardless of its necessary cohesive strength, which is compatible with that of rubble pile asteroids, it may preferably have a monolithic structure \citep{hirabayashi2021hayabusa2, santana2025hayabusa2}. Monolithic asteroids can exhibit thermal conductivities an order of magnitude greater than those of rubble piles \citep{luan2015thermal}. Additionally, sublimation on rubble piles can leave behind a dehydrated subsurface layer that shields against further ice losses \citep{zhang2024characterizing}. For these reasons, outgassing rates of deeply buried water is believed to be relatively high on monoliths.}

The present work investigates the thermal conditions in 1998 KY26's interior under its current orbital conditions and examines whether volatiles can exist in a stable state. Of the compounds commonly designated as volatiles in the discussion of comets \citep{a2012cometary}, water has the highest sublimation temperature. Since an anhydrous asteroid would likely be depleted in other volatile species, this work focuses its discussion on the presence of water. We employ the thermophysical finite element model (TPFEM), originally developed by \cite{nakano2023finite}. TPFEM is modified for this study to examine the thermal behavior deep in the subsurface and is used to numerically characterize the temperatures of nodes in the asteroid's shape model in a quasi-steady state. TPFEM takes orbital position and intensive thermophysical properties as inputs, allowing for temperature information to be averaged over the asteroid's entire orbit. The impact of variations in thermophysical properties on temperatures is also characterized to account for uncertainties in current observations of 1998 KY26.

This study is organized as follows. In Section \ref{sec:Methodology}, we discuss the principles of TPFEM and its subsequent modifications. To complete the characterization of the simulation setup, input parameters for TPFEM---orbital positions, simulation time, and thermophysical parameters---are enumerated, and a control set (base case) is defined. Section \ref{sec:Results} presents spatial temperature variation results for all test cases. Deviations from the base case bulk average temperature are also listed to illustrate the relative impact of adjustments to the input parameters. The implications of this temperature data on 1998 KY26's water content are discussed in Section \ref{sec:Discussion}, followed by concluding remarks in Section \ref{sec:Conclusion}.

\section{Methodology}
\label{sec:Methodology}

\subsection{Thermophysical characterization basis}

\textcolor{black}{TPFEM was selected for this study for its availability, accuracy, and validation against in-situ spacecraft observations \citep{nakano2023finite}. Thermophysical models such as this account for local shadowing, sunlight scattering, and three-dimensional (3D) heat conduction \citep{nakano2025modeling}, making them useful for modeling irregular bodies such as asteroids. Lateral heat conduction becomes critical at thermal equilibrium, particularly in deeper subsurface layers, as the heat budget balances in 3D. Thus, though this study uses a coarse shape model, 3D conduction is still important for the present problem.} TPFEM solves the 3D heat conduction equation using the Runge-Kutta (RK4) method \citep{nakano2023finite}. The governing equation in integral form is given as follows:
\begin{equation}
    \iiint_V \rho c_p \frac{\partial T}{\partial t} dV = -\iint_S \bfm{q} \cdot d\bfm{S} - \iint_S \bfm{Q} \cdot d\bfm{S}
    \label{eq:TPFEM1}
\end{equation}
where $T$ is temperature, $V$ is volume, $S$ is (surface) area, $\rho$ is density, $c_p$ is specific heat capacity, $\bfm{q}$ is the heat flux at the surface boundary, and $\bfm{Q}$ is the heat energy at the boundary. $\bfm{q}$ can be described as a diffusivity $\kappa$ (thermal conductivity) times a thermal gradient based on Fourier's thermal law. \textcolor{black}{Given Galerkin's method \citep{ern2004theory, nakano2025modeling}} and four-node tetrahedral mesh elements, Equation (\ref{eq:TPFEM1}) is transformed into Equation (\ref{eq:TPFEM2}) for one tetrahedral mesh element:
\begin{equation}
    \bfm{A}^e \dot{\bfm{T}^e} + \bfm{B}^e \bfm{T}^e + \bfm{C}^e = 0
    \label{eq:TPFEM2}
\end{equation}
where the temperature at each element ($\bfm{T}^e$) is a column vector of the temperatures of its four nodes. $\bfm{A}^e$ and $\bfm{B}^e$ are 4 by 4 matrices per mesh element. \textcolor{black}{Throughout the present study, $\rho$, $c_p$, and $\kappa$ are set constant, which is done for two reasons. First, this prevents $\bfm{A}^e$ and $\bfm{B}^e$ from being continually reevaluated, increasing computational efficiency. Second, the temperature response of materials on 1998 KY26 is not well characterized. Certain studies of inner solar system objects have expressed $c_p$ and $\kappa$ as functions of temperature to account for diurnal temperature variations \citep{hayne2017global}. The impact of incorporating temperature-dependent $c_p$ and $\kappa$ here would be to have those parameters reach values different from what was inputted. In Section \ref{sec:Results}, we show that both $c_p$ and $\kappa$ have a limited effect on bulk temperature, despite being varied over a large range. Thus, we expect our results to hold even if these parameters are made temperature dependent.}

\begin{figure}
\centering
\includegraphics[scale=0.50]{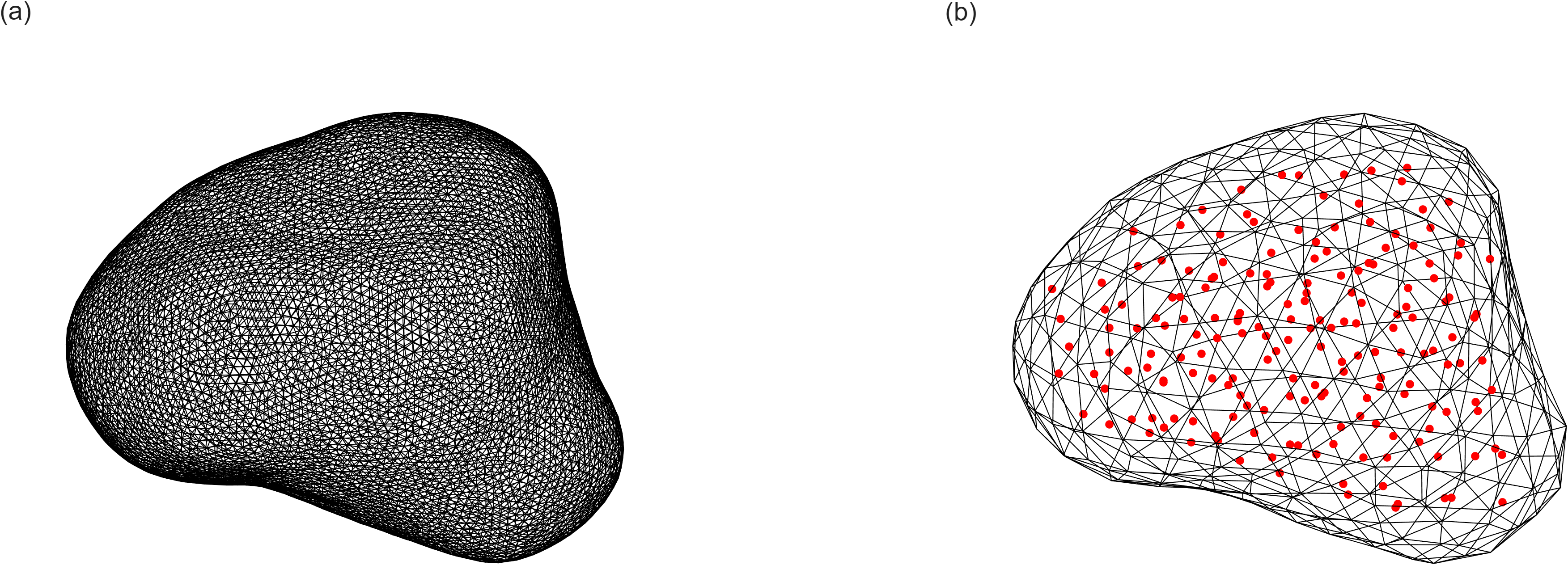}
\caption{The 1998 KY26 shape model (a) in its original form and (b) its working form after quadric edge collapse decimation. Interior nodes produced by Tetgen are highlighted for the simplified model.}
\label{fig:shapeModelSurf}
\end{figure}

Unlike the earlier study that focused on the temperature variations at a similar scale to the thermal skin depth of asteroids \citep{nakano2023finite}, we use TPFEM to probe the temperature deeper into the target asteroid's center. Following earlier finite element modeling studies \citep{hirabayashi2021finite}, we start with a polyhedral shape model consisting of triangular facets. \textcolor{black}{Tetgen software \citep{si2013tetgen} is then used to generate a minimal quantity of internal nodes and volume elements using Delaunay tetrahedralization. We apply this scheme to the concave shape model derived using light curve data by \cite{santana2025hayabusa2}. The original shape parameters are 8571 vertices and 17138 facets. Because the cost of computing Equation (\ref{eq:TPFEM2}) increases quadratically with element count, we reduced the vertex and facet sizes to 276 vertices and 548 facets using the quadric edge collapse decimation method in MeshLab \citep{cignoni2008meshlab}. These working parameters were chosen such that simulations could be completed in approximately one day. This process expanded the average facet size from 0.024 m$^2$ to 0.739 m$^2$, which is visualized in Figure \ref{fig:shapeModelSurf}. The validation procedure for this 
simplified model, including commentary on shadowing conditions, is provided in Appendix \ref{app:SMV}. Although resolution is reduced, computational efficiency is prioritized because both thermophysical inputs and orbital position are varied, meaning 260 simulations are run in total.} At the end of each simulation, the bulk average temperature of the asteroid is calculated, where element temperatures are summed and weighted by volume.

\subsection{Numerical simulation settings}
\label{sec:NumericalSettings}

Our numerical analysis requires heliocentric positions, simulation time, and thermophysical parameters as inputs. Thermophysical parameters are varied, whereas a standardized simulation time and set of orbital positions are used. We conduct 13 tests to assess the influence of spin axis orientation, size (equivalent diameter), thermal conductivity, bond albedo, and (constant-pressure) specific heat capacity on the temperature of 1998 KY26. The first test is the base case, which contains the control values for the thermophysical parameters. Each parameter is then varied in isolation, with a maximum and minimum value equally spaced from the base case value.

For each thermophysical parameter variation, 20 heliocentric positions are simulated. These positions correspond to $\nu=18\degree$ true anomaly intervals in a fixed orbital state, where the asteroid model rotates in place. The bulk average temperature data from each position is subsequently time-averaged over the asteroid's orbit, so simulations with true anomalies closer to apoapsis carry a greater weight. This orbital state is defined using ephemeris information over 425 years downloaded from the JPL Horizons database. The ephemeris is converted into one set of osculating orbital elements per month, and then averaged. The working orbital elements for this study are shown in Table \ref{tab:simulationInputs}. At each heliocentric position, TPFEM requires components for the Sun vector, which points towards the asteroid in the inertial heliocentric ecliptic (EJ2000) frame. When it runs, TPFEM performs a 3-2-3 Eulerian rotation about ecliptic longitude ($\lambda$), ecliptic latitude ($\beta$), and rotational phase to obtain the Sun vector in the body-fixed frame \citep*{DAMIT2024} at any simulated time. \textcolor{black}{Due to the small (10 second) timestep used to model the thermophysics of this fast-rotating asteroid, the Sun vector is not continuously rotated. While orbits are not continuous, the resolution afforded by dividing the orbit into 20 arcs was deemed sufficient. The asteroid's bulk average temperature over a full orbit decreased by $0.02\%$ when the resolution was doubled (40 positions used) and increased by $0.09\%$ when the resolution was halved (10 positions used). These tests show that changing the orbit resolution has a negligible impact on temperature, and that the magnitude of temperature differences decreases as resolution increases.}

Simulation time must be set so that each run can settle from the model's initial condition. To ensure the TPFEM simulation is long enough to make the thermal environment reach a quasi-steady state, where the temperature variation only shows a periodic mode (from the asteroid's day-night cycle), the time to reach such a state is conservatively estimated. A simple analytical solution using a spherical target as a simple model \citep{bergman2011introduction} yields a settlement period of 613 Earth days. Given this settlement process, we confirm that, for all cases, the bulk average temperature from the first simulated 1998 KY26 day (5.35 minutes) deviates from the last simulated day (24 hours later) by less than $0.001\%$. This shows that a quasi-steady state is reached, so this simulation time is used for all cases in this study.

When it comes to selecting parameter settings for the base case, remote observational data carry a large uncertainty, particularly for thermal properties. It is thus assumed, when necessary, that 1998 KY26 is a monolithic E-type asteroid, which is a favorable, but not guaranteed interpretation \citep{kikuchi2023preliminary}. The enstatite series presented by \cite{krupka1985low} informs the base case $c_p$. $c_p$ is temperature-dependent, so a bulk temperature of 229.740 K is used. \textcolor{black}{This value is determined by initially using the $c_p$ value of a validated thermophysical study of the asteroid Bennu: 750 J/kg$\cdot$K \citep{nakano2023finite}. The $c_p$ of enstatite at the temperature output of this first seed simulation is used to set up a second seed simulation, which yields 229.740 K. The base case $c_p$ is thus set at 670.6 J/kg$\cdot$K.} The minimum value is the original seed $c_p$ of 502.9 J/kg$\cdot$K, while the maximum value is 838.2 J/kg$\cdot$K (\textcolor{black}{Table \ref{tab:simulationInputs}}). \cite{maclennan2022thermophysical} give expected values and uncertainties for thermal conductivity and thermal emissivity for E-type meteorites. Thermal conductivity, like $c_p$, is presented as a (weak) function of temperature. Using the same seed temperature as for $c_p$, a base case $\kappa$ of 5.403 W/m$\cdot$K is calculated, with a minimum value of 4.863 W/m$\cdot$K and a maximum value of 5.943 W/m$\cdot$K.

The remaining parameters are picked using observations of 1998 KY26 specifically. \cite{santana2025hayabusa2} estimated the asteroid's spin pole orientation, equivalent diameter, and geometric albedo, all with their respective uncertainties. Since spin pole orientation is a tuple ($\lambda$, $\beta$), it requires 4 variations instead of the usual 2. These variations cover the corners of the asteroid's candidate pole region: ($30\degree$, $-60\degree$), ($45\degree$, $-60\degree$), ($30\degree$, $-30\degree$), and ($45\degree$, $-30\degree$). \cite{santana2025hayabusa2} derive ($36\degree$, $-44\degree$) as the asteroid's spin pole orientation, which is used in the base case here. Equivalent diameter, discussed above, is set at 11 m for the base case, and bulk density is set at 2800 kg/m$^3$ for the base case, following the earlier astrometry analysis \citep{farnocchia2025radiation}.

Observations report geometric albedo ($p_V$) for 1998 KY26, but TPFEM requires Bond albedo ($A_B$), the full-spectrum reflectance ratio, as an input. $A_B$ is equal to $p_V$ multiplied by the asteroid's phase integral, which \cite{mainzer2015space} define in terms of its magnitude slope $G$. The Small Body Database \citep{SBDB1998KY26} compiles observations from \cite{santana2025hayabusa2} to give $G=0.42\pm0.02$. Both the uncertainty in $G$ and the $p_V$ are considered when setting bounds for $A_B$ in Table \ref{tab:simulationInputs}.

\begin{table}[width=.9\linewidth,cols=6,pos=ht]
\caption{Geophysical, thermal, and orbital settings used in TPFEM simulations. The base case contains the control or median values of all parameters. Certain settings may be individually varied from the base case to the listed maximum and minimum values.}
\label{tab:simulationInputs}
\begin{tabular*}{\tblwidth}{@{} LLLLLL@{} }
\toprule
Parameter & Symbol & Units & Base Value & Minimum Value & Maximum Value \\
\midrule
Ecliptic longitude & $\lambda$ & $\degree$ & 36 & 30 & 45\\
Ecliptic latitude & $\beta$ & $\degree$ & -44 & -60 & -30\\
Equivalent diameter & $d_{eq}$ & m & 11 & 9 & 13\\
Specific heat capacity & $c_p$ & J/kg$\cdot$K & 670.6 & 502.9 & 838.2\\
Thermal conductivity & $\kappa$ & W/m$\cdot$K & 5.403 & 4.863 & 5.943\\
Bond albedo & $A_B$  & --- & 0.300 & 0.245 & 0.355\\
Bulk density & $\rho$ & kg/m$^3$ & 2800 & --- & ---\\
Thermal emissivity & $\epsilon$ & --- & 0.9 & --- & ---\\
Semi-major axis & $a$ & au & 1.26 & --- & ---\\
Eccentricity & $e$ & --- & 0.21 & --- & ---\\
Inclination & $i$ & $\degree$ & 1.04 & --- & ---\\
Argument of ascending node & $\Omega$ & $\degree$  & 105.86 & --- & ---\\
Argument of periapsis & $\omega$ & $\degree$  & 183.39 & --- & ---\\
\bottomrule
\end{tabular*}
\end{table}

In sum, the strategy for thermophysical parameter variation is as follows: we present a base case set of parameters and assumptions that forms a reasonable estimate for the conditions on 1998 KY26, then, due to observational uncertainties, we vary the parameters one at a time to isolate their impact on bulk temperature. We use the base case estimate the water loss rate. If the impact of certain parameters on temperature is minuscule, we can be confident in our predictions despite existing uncertainties.

\subsection{Water loss rate estimation}
\textcolor{black}{We use a uniform bulk temperature to estimate a water loss rate ($\dot{m}_{H_2O}$, kg/s$^2$) by applying the H$_2$O flux ($\Bar{J}$) relation for buried ice presented by \cite{schorghofer2008lifetime}:
\begin{equation}
    \Bar{J} = \frac{2 \pi}{8 + \pi}\frac{\phi}{1-\phi}\frac{1}{\tau}\sqrt{\frac{2 m_{H_2O}}{\pi k_B T}}\frac{R_g}{z}p_s
    \label{eq:SchorghoferWater}
\end{equation}
where $\phi$ is bulk porosity, $\tau$ is tortuosity, $m_{H_2O}$ is the mass of a water molecule, $k_B$ is Boltzmann's constant, $R_g$ is grain radius, $z$ is burial depth, and $p_s$ is equilibrium vapor pressure for H$_2$O. All values in Equation (\ref{eq:SchorghoferWater}) are held constant except for temperature and $p_s$. Using the E-type grain density of 3150 kg/m$^3$ given by \cite{maclennan2022thermophysical} and the bulk density from Table \ref{tab:simulationInputs}, porosity is set to 0.111. Thus, using the Bruggeman sphere relation \citep{holzer2023tortuosity}, tortuosity is 3. Burial depth is set to 5.5 m---the asteroid's equivalent radius---to make our water loss rate estimates conservative. Similarly, grain radius is set to 32 $\mu$m, the lower bound of what was derived for 1950 DA \citep{gundlach2015regolith}, a better characterized asteroid with high albedo. Using the vapor pressure of water ice obtained by \cite{murphy2005review} makes Equation (\ref{eq:SchorghoferWater}) a function of temperature, $T$, for our purposes. Lastly, $\Bar{J}$ is multiplied by the shape model surface area of 404.9 m$^2$ to obtain $\dot{m}_{H_2O}$.}

\section{Results}
\label{sec:Results}

\begin{figure}
\centering
\includegraphics[scale=0.73]{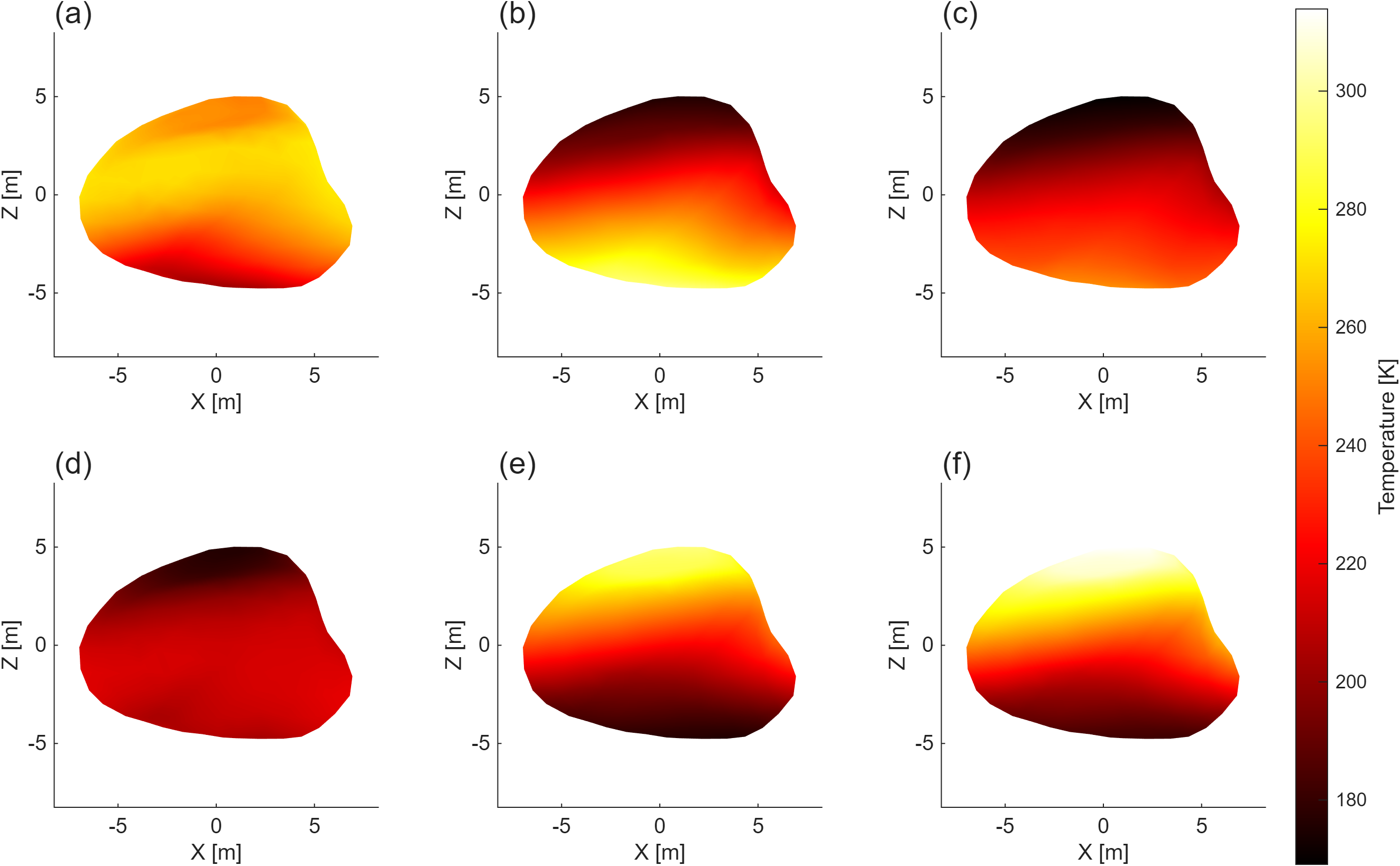}
\caption{Surface temperature profile for 1998 KY26. Panels (a) through (c) show $\nu=0\degree$ (periapsis), $\nu=90\degree$, and $\nu=144\degree$, where the coldest recorded surface node occurs. On the other hand, Panels (d) through (f) illustrate $\nu=180\degree$ (apoapsis), $\nu=270\degree$, and $\nu=306\degree$, where the hottest recorded surface node occurs.}
\label{fig:CaseTempExamples}
\end{figure}

Each of the 20 simulations per test produces different nodal temperature distributions as the orbital position changes, as shown in Figure \ref{fig:CaseTempExamples}. While the changing of the asteroid's seasons may create transient cold patches where water ice can exist, \textcolor{black}{water cannot exist long-term at any given location if its temperature exceeds 145 K, which is a well-used sublimation temperature threshold for buried water \citep{schorghofer2008lifetime}. The sublimation threshold without burying is commonly 110 K \citep{williams2019seasonal}. We present orbit-averaged temperatures for the purpose of readability, as any node with an orbit-averaged temperature above 145 K has inherently crossed the water sublimation threshold in at least one instance in time.} After the node average temperature array is compiled, nodes are sorted by radial distance from the model center into 30 bins so a standard deviation ($1\sigma$) spread can be plotted.

The long-term bulk average temperature of 1998 KY26 is computed to be 225.737 K for the base case. This value far exceeds the sublimation temperature for water ice in a vacuum, 145 K, suggesting that volatile existence at present is largely unstable. To see how local---not bulk---temperatures change with location and time, Figure \ref{fig:CaseTempExamples} is useful. While the asteroid is coldest on average at apoapsis and hottest on average at periapsis, these are not the positions of extreme local temperatures. At $\nu=144\degree$, \textcolor{black}{the northern hemisphere 1998 KY26 cools to 169.170 K}. At $\nu=306\degree$, however, the same region reaches 309.191 K, showing that there are no perennial surface cold patches on 1998 KY26.

\begin{figure}
\centering
\includegraphics[scale=0.61]{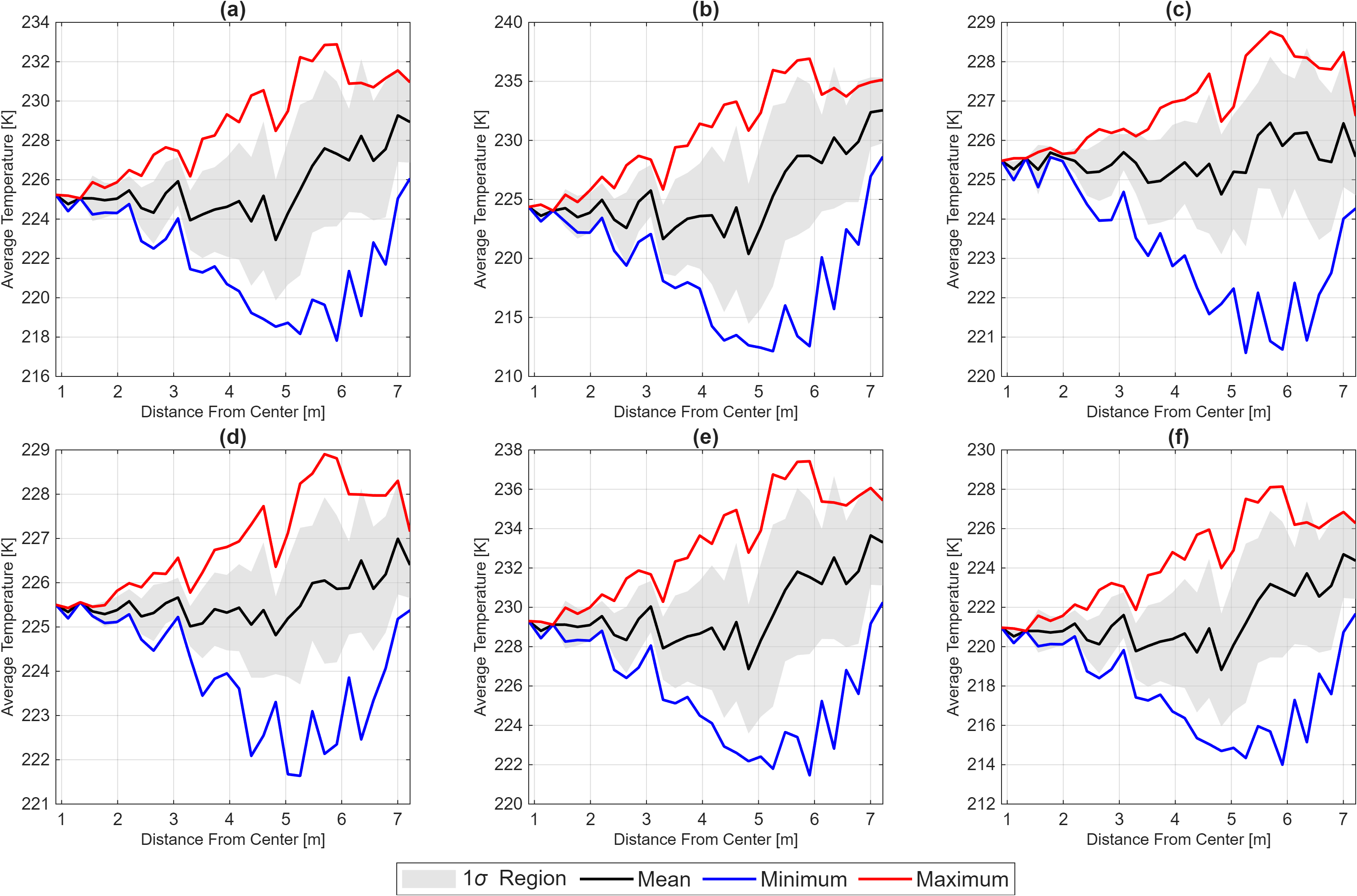}
\caption{Time-averaged node temperature variation with radial distance. For the upper panels, Panels (a) through(c) show the base case,  $\lambda=45\degree$ \& $\beta=-60\degree$, and $\lambda=30\degree$ \& $\beta=-30\degree$. For the lower panels, Panels (d) through (f) display $\lambda=45\degree$ \& $\beta=-30\degree$, $A_B=0.245$, and $A_B=0.355$.}
\label{fig:tempVariation}
\end{figure}

While Figure \ref{fig:CaseTempExamples} only provides data about the surface of 1998 KY26, Figure \ref{fig:tempVariation} shows the time-average temperatures of interior nodes. The shapes of these plots align with expectations---temperature variation is least near the asteroid's core and greatest near the equivalent radius (5.5 m), which is about where a majority of the asteroid's surface area resides. Points with a radial distance greater than 5.5 m represent the peaks on 1998 KY26's irregular surface. Because significant shadowing effects are not expected in the peaks region, the temperature variation tapers towards the mean above 5.5 m.

Ultimately, while the core region is approximately 1 K colder than the surface for the base case, it is very close to the bulk average temperature. Figure \ref{fig:tempVariation} identifies long-term outlying cold nodes at any depth, with the minimum temperature being slightly below 218 K for the base case. Since this node is further than 5.5 m from the model center, this represents a shadowed region on the surface. All subsurface nodes are hotter. The narrow variance shown in Figure \ref{fig:tempVariation} is consistent among all test formulations. The minimum orbital-averaged node temperature occurs during the $\lambda=45\degree$, $\beta=-60\degree$ test case, which is approximately 212 K. 

\begin{table}[width=.9\linewidth,cols=5,pos=ht]
\caption{Bulk average temperatures for thermophysical property variations. In each case, the remaining parameters apply those in the base case.}
\label{tab:bulkMeanTemps}
\begin{tabular*}{\tblwidth}{@{} LLLLL@{} }
\toprule
Test ID & Parameter(s) varied & Value(s) & Bulk Average Temperature (K) & Deviation from Base (K)\\
\midrule
Base Case & --- & --- & 225.737 & ---\\
1 & $\lambda$, $\beta$ & $30\degree, -60\degree$ & 225.484 & -0.253\\
2 & $\lambda$, $\beta$ & $45\degree, -60\degree$ & 225.489 & -0.248\\
3 & $\lambda$, $\beta$ & $30\degree, -30\degree$ & 225.583 & -0.154\\
4 & $\lambda$, $\beta$ & $45\degree, -30\degree$ & 225.609 & -0.128\\
5 & $d_{eq}$ & $9$ m & 226.407 & 0.670\\
6 & $d_{eq}$ & $13$ m & 225.146 & -0.591\\
7 & $\kappa$ & $4.863$ W/m$\cdot$K & 225.367 & -0.370\\
8 & $\kappa$ & $5.943$ W/m$\cdot$K & 226.061 & 0.324\\
9 & $A_B$ & $0.245$ & 229.848 & 4.111\\
10 & $A_B$ & $0.355$ & 221.433 & -4.304\\
11 & $c_p$ & $502.9$ J/kg$\cdot$K & 225.737 & 0.000\\
12 & $c_p$ & $838.2$ J/kg$\cdot$K & 225.737 & 0.000\\
\bottomrule
\end{tabular*}
\end{table}

Of the parameters varied in Table \ref{tab:bulkMeanTemps}, albedo had by far the strongest impact on temperature, while spin pole axis had the least observable impact. \textcolor{black}{Changing the specific heat capacity, despite constituting the greatest percentage variation of an input ($25\%$), had no visible effect on the bulk average temperature.} This is not to say, however, that the outputs of the $c_p$ tests are identical to the base case; examining the raw TPFEM node temperature outputs shows that day-night temperature amplitude increases by 31.71\% at periapsis and by 44.44\% at apoapsis for the low $c_p$ case; likewise, the high $c_p$ case corresponds to a 21.43\% drop in day-night temperature amplitude at periapsis and a 22.22\% drop at apoapsis. Though not presented in Table \ref{tab:bulkMeanTemps}, varying the asteroid's bulk density has a similar effect to varying $c_p$, as neither parameter appears independently of the other in the equations that underpin TPFEM. \textcolor{black}{While thermal inertia ($\Gamma$) is a common thermophysical parameter, it is also not shown in Table \ref{tab:bulkMeanTemps}, as it is a compound parameter of $\kappa$, $c_p$ and $\rho$. The highest and lowest $\Gamma$ values correspond to the high and low $c_p$ cases---3561.0 and 2758.3 tiu, respectively. While these values exceed what has been observed by spacecraft at rubble pile asteroids \citep{okada2020highly}, this is attributed to the high conductivity and density parameters stemming from the previously established monolithic structure interpretation. These $\Gamma$ values are comparable to similarly monolithic meteorites \citep{opeil2020surprising}. The $\Gamma$ values for the high and low $\kappa$ cases are 3340.5 and 3021.8 tiu, demonstrating that changing $\Gamma$ does not have a measurable effect on bulk temperature outside of what was observed by varying conductivity.} All told, no variations in thermophysical parameters within their respective predicted uncertainties alter the conclusions drawn from the base case.

\begin{table}[width=.9\linewidth,cols=6,pos=ht]
\caption{Case temperature variations with orbital positions.}
\label{tab:caseTempsOrbital} 
\begin{tabular*}{\tblwidth}{@{} LLLLLL@{} }
\toprule
\shortstack{Test ID} & \shortstack{Bulk Temperature \\ at Apoapsis (K)} & \shortstack{Bulk Temperature \\ at Periapsis (K)} & \shortstack{Coldest Surface Node \\ Temperature (K)} & \shortstack{Hottest Surface Node \\ Temperature (K)} & \shortstack{$\nu$ for Coldest \\ Surface Node} \\
\midrule
Base Case & 205.035 & 253.198 & 169.104 & 313.908 & $144\degree$\\
1 & 204.867 & 253.145 & 174.062 & 285.625 & $144\degree$\\
2 & 205.057 & 253.220 & 172.147 & 291.052 & $144\degree$\\
3 & 204.932 & 253.153 & 166.575 & 324.840 & $108\degree$\\
4 & 205.833 & 253.301 & 164.167 & 331.744 & $126\degree$\\
5 & 205.272 & 253.600 & 173.561 & 311.694 & $144\degree$\\
6 & 204.815 & 252.847 & 165.023 & 315.518 & $126\degree$\\
7 & 204.898 & 252.979 & 166.560 & 314.947 & $126\degree$\\
8 & 205.151 & 253.392 & 171.238 & 312.895 & $144\degree$\\
9  & 208.877 & 257.917 & 170.971 & 320.499 & $126\degree$\\
10 & 201.015 & 248.263 & 167.058 & 307.005 & $144\degree$\\
11 & 205.035 & 253.198 & 169.104 & 313.909 & $144\degree$\\
12 & 205.035 & 253.198 & 169.104 & 313.907 & $144\degree$\\
\bottomrule
\end{tabular*}
\end{table}

Building upon these results, Table \ref{tab:caseTempsOrbital} shows that the test variations did not lead to large deviations from the base case when bulk temperature is measured specifically at apoapsis or periapsis, indicating that temperature deviations observed in Table \ref{tab:bulkMeanTemps} are experienced roughly uniformly over an orbit. Variations due to albedo again stand out as the most significant. Table \ref{tab:caseTempsOrbital} also shows that the most extreme cold and hot surface nodes are more sensitive to the test variations compared to bulk temperatures, which is expected. Spin pole axis variations, despite leading to small deviations in bulk average temperature compared to all thermophysical parameters (except $c_p$), have a comparable if not greater impact on minimum temperature. The true anomaly associated with this coldest surface temperature is shown to vary by up to $36\degree$, whereas the true anomaly associated with the hottest surface temperature is consistently $306\degree$.

\section{Discussion}
\label{sec:Discussion}

The dark comet concept is a proposed class of small bodies that explains why astrometric studies on certain asteroids yield significant nongravitational accelerations that are difficult to resolve with \textcolor{black}{solar pressure or thermal emission forces}: the asteroids are perturbed by outgassing \citep{seligman2024two}. The water loss rates on dark comets are implied from the magnitude of this acceleration, which, if they are to be sustained long-term, imply a significant presence of volatiles. The dynamically implied water loss rate on 1998 KY26 is multiple orders of magnitude lower than that of comets \citep{seligman2023dark}, suggesting that the asteroid's nongravitational acceleration is sustainable. The hypothesis that 1998 KY26 is a dark comet can be challenged if its thermally calculated volatile loss rate is too high to be sustained over its lifetime. Since volatile sublimation rates are a function of vapor pressure, which in turn depends on temperature \citep{schorghofer2008lifetime}, our results are well suited to this discussion. Ultimately, we contend that 1998 KY26 is dehydrated on two main points: (1) the asteroid loses volatiles too rapidly because its interior is too hot, and (2) its temperature remains consistent despite significant uncertainty in thermophysical data. This discussion is limited to 1998 KY26 specifically, and does not cover the general applicability of the dark comet concept.

Our results showed that the bulk average temperature of 1998 KY26 remained largely insensitive to all parameters varied in this study, as shown in Table \ref{tab:bulkMeanTemps}. Furthermore, variations in thermophysical parameters had little discernible impact on the spatial distribution of hot and cold zones, as shown in Figure \ref{fig:tempVariation}. These figures show that node temperatures remain above the sublimation threshold regardless of the \textcolor{black}{burial depth or shadowing conditions}. The lack of any long-term cold traps on this asteroid aligns with studies linking their extent to spin pole orientation. High obliquities, like those exhibited by all test cases here, are shown to make cold traps nearly nonexistent \citep{siegler2015evolution, schorghofer2024history}. The asteroid's small size also significantly contributes to the lack of intense shadowing, which is demonstrated in Table \ref{tab:caseTempsOrbital}, where the 13 m equivalent diameter test case yields the coldest minimum temperature. In sum, all test cases had a consistent bulk average temperature and a lack of exceptional shadowing. \textcolor{black}{The insensitivity of temperature to any of the parameters varied is further demonstrated in Appendix \ref{app:RubblePile}, where thermal conductivity befitting a rubble pile structure is assumed, resulting in a bulk temperature of 212.428 K---still well above the sublimation threshold.} Thus, further calculations can be simplified by only considering the base case and applying a uniform temperature.

\textcolor{black}{We translate our bulk temperature result for the base case into a water loss rate using Equation (\ref{eq:SchorghoferWater}).} The resultant $\dot{m}_{H_2O}$ is $3.345\times10^{-7}$ kg/s$^2$, sufficient to sublimate an initial water content equivalent to 1\% of the asteroid's mass in 2065 years. This initial water content is well above the 0.54\% upper bound for enstatite chondrites \citep{piani2020earth}, which themselves contain more water than enstatite achondrites (E-types). Furthermore, not all initial water would be buried at the asteroid's core, so this timescale is still an overestimate. If our sublimation rate is increased to the outgassing rate derived from nongravitational acceleration magnitude, which is $\sim$10$^{-5}$ kg/s$^2$ \citep{bolin2025keck}, the time to fully dehydrate the asteroid drops to 62 years. Both of these time estimates are almost certainly below the current age of the asteroid, which can be deduced by examining its orbit. If the asteroid had a volatile-rich parent body, it would likely reside in the main belt; the Hungaria \citep{santana2025hayabusa2} or Nysa \citep{tatsumi2026hayabusa2} families, specifically, have been cited as plausible origins. 1998 KY26, however, currently resides in a near-Earth orbit. \cite{bottke2002origin} shows that the timescale for main belt asteroids below the $\nu_6$ resonance to evolve into Earth-crossers is on the order of 60 million years. Asteroids further into the main belt would have to rely on less powerful transport mechanisms (i.e., mean motion resonances, Yarkovsky effect) and can be generally assumed to take even longer to evolve into Earth-crossers. All told, it is exceedingly unlikely that 1998 KY26 is young enough to retain water. 

The temperature data presented in this study can be reconciled with the anomalous $A_3$ nongravitational acceleration term observed in 1998 KY26 by \cite{seligman2023dark}, which sparked discussion of the asteroid as a dark comet with water production. \cite{farnocchia2025radiation} \textcolor{black}{analytically models the combination of direct solar radiation pressure (absorption and emission) and thermal emission---which includes, but is not limited to, the Yarkovsky effect \citep{vokrouhlicky2015yarkovsky}---on an oblate spheroid, relating 1998 KY26's nongravitational acceleration terms} to the asteroid's spin pole axis orientation. While \textcolor{black}{these forces have previously been relegated to the discussion radial and transverse accelerations, asteroids with sufficient obliquity can exhibit normal (out-of-plane) accelerations as well. This is demonstrated in the case of 1998 KY26, where} \cite{farnocchia2025radiation} achieves a good fit to orbital data with the best-fit spin pole axis orientation---$\lambda=37.7\degree$, $\beta=-56.4\degree$---being within the candidate spin pole region defined by \cite{santana2025hayabusa2}. As our study proposes no sublimation-driven forces acting on 1998 KY26, we also support this astrometric study, which suggests solar radiation pressure \textcolor{black}{and thermal emission} contributing to out-of-plane nongravitational effects. 

\section{Conclusion}
\label{sec:Conclusion}

Using the thermophysical finite element modeling (TPFEM) approach, we determined that 1998 KY26 is \textcolor{black}{likely a dehydrated asteroid.} We performed a simulation using measured estimates of the spin pole-axis orientation and thermophysical parameters, and showed an orbit-averaged bulk temperature far above the sublimation threshold for ice in a vacuum. Our post-simulation analysis showed that the spatial variation of temperature was minimal---both between the average surface and core temperatures, and between the hottest and coldest surface nodes. Thus, the long-term presence of exceptionally shadowed regions was ruled out. Temperature variations from changing thermophysical parameters were also minor (with the possible exception of albedo), suggesting that these conclusions are consistent despite significant observational uncertainty. Using this temperature information, we calculated that the water loss rate on 1998 KY26 is too high for it to retain volatiles in the present, endorsing the hypothesis that the asteroid's out-of-plane acceleration is driven by solar radiation forces.

\appendix
\section{Simplified shape model validation}
\label{app:SMV}

\textcolor{black}{Attempting to run TPFEM with the original number of vertices (see Figure \ref{fig:shapeModelSurf}) pushes the tetrahedral element count over an operational limit, so the temperature results from the working (simplified) shape model cannot be directly compared to the unmodified model. However, we show that despite significant increases in element count, bulk temperatures remain consistent, meaning our results from simulating a simplified model can be extrapolated to much more detailed shapes.}

\textcolor{black}{Two parameters control the number of tetrahedral elements of the models inputted to TPFEM: target surface vertex count in Meshlab simplification and maximum element volume allowed in Tetgen. Thus, we compare our working shape model to a model with the same number of surface vertices but with a 4x tighter maximum element volume (Model 2), a model with the same element volume maximum but with 4x the surface vertices (Model 3), and a model with a 4x tighter maximum element volume and 4x the surface vertices (Model 4). We compare the bulk average temperature of each model after 12.5 simulated days, chosen so Model 4 uses the same CPU time as the working model uses in the rest of this study. Other TPFEM inputs are those of the base case at periapsis, as defined in Section \ref{sec:NumericalSettings}.}

\begin{figure}
\centering
\includegraphics[scale=0.46]{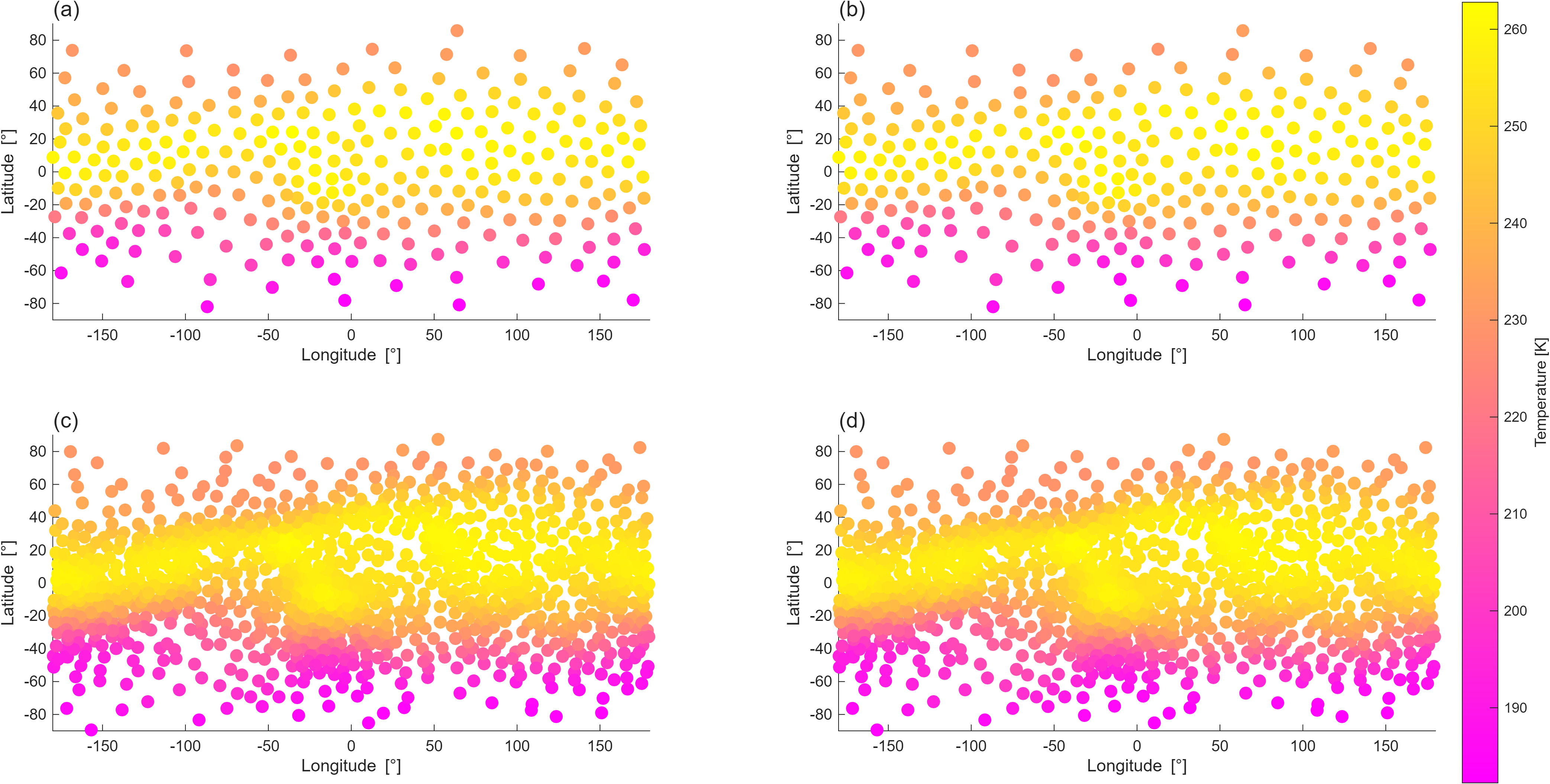}
\caption{Surface node temperature distribution after 12.5 days for (a) the working shape model, (b) a shape model with 4x smaller maximum tetrahedral element size, (c) a shape model with 4x the surface nodes, (d) a shape model with both refinements applied.}
\label{fig:ShadowingSurf}
\end{figure}

\textcolor{black}{In this comparison, the working shape model reaches a bulk temperature of 225.899 K. Model 2 reaches 225.811 K, Model 3 reaches 225.832 K, and Model 4 reaches 225.804 K. This constitutes at most a $0.04\%$ reduction in temperature, though it should be noted that this comparison was not run at a quasi-steady state (defined in Section \ref{sec:NumericalSettings}). As the working model reaches a steady temperature of 253.198 K, one can expect the temperature difference to approximately double. Regardless, the bulk temperatures of all models are highly consistent. Furthermore, it is shown in Figure \ref{fig:ShadowingSurf} that none of the more detailed models uncover any new exceptionally cold or hot surface nodes nor qualitatively change their distribution. In sum, statements made about bulk temperature or shadowing conditions on the working model can be extended to more detailed models.}

\section{Temperature computation for a rubble pile structure}
\label{app:RubblePile}

\textcolor{black}{While 1998 KY26 having a monolithic structure is preferred, measurements of the strength of the Yarkovsky effect on the asteroid may suggest existence of a surface layer of regolith \citep{petkovic2021extended}. Rubble piles have drastically lower thermal conductivities than monoliths, which may insulate buried water. To rule out this possibility, we model the entire asteroid as having rubble pile conductivity. The value of $\kappa$ we use in this simulation, 0.192 W/m$\cdot$K, is obtained by assuming the asteroid has a thermal inertia on par with the rubble pile Ryugu: 600 tiu \citep{okada2020highly}. Heat capacity and density are kept at their base case values.}

\begin{figure}
\centering
\includegraphics[scale=0.5]{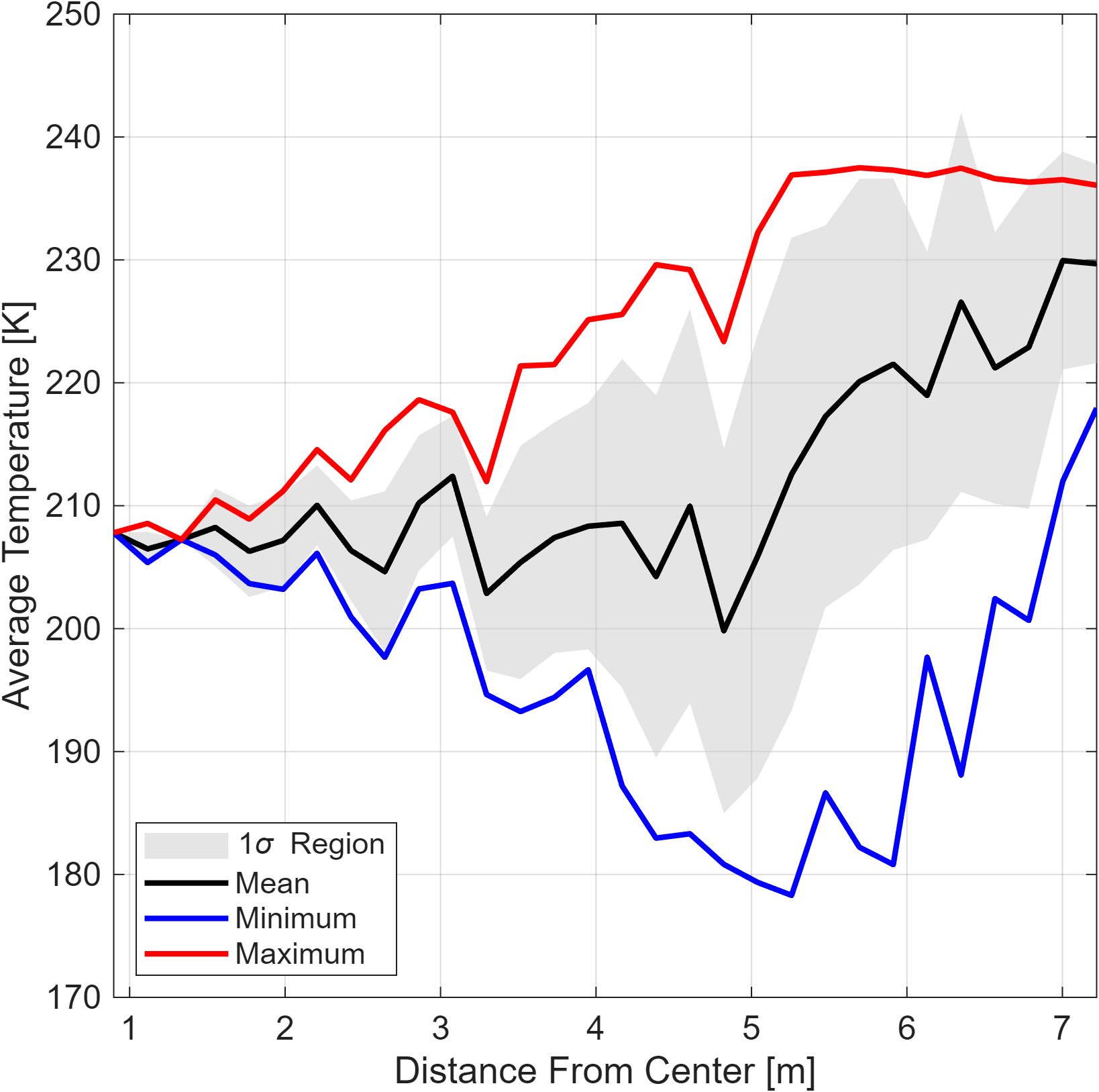}
\caption{Time-averaged node temperature variation with radial distance for $\kappa=0.192$ W/m$\cdot$K.}
\label{fig:RPTempVariation}
\end{figure}

\textcolor{black}{Despite $\kappa$ being reduced by over an order of magnitude relative to the base case, the orbit-averaged bulk temperature comes out to 212.428 K---reduced, but still in the same regime as all other test cases. Figure \ref{fig:RPTempVariation} shows that while the variance in node temperatures is increased, no nodes cross the 145 K sublimation threshold. These results show that our results hold even if the assumption of a monolithic structure does not.}

\printcredits

\section*{Declaration of competing interest}
The authors declare that they have no known competing financial interests or personal relationships that could have appeared to influence the work reported in this paper.

\section*{Acknowledgments}
The study is supported by the Georgia Institute of Technology. We also acknowledge the support in part through research cyberinfrastructure resources and services provided by the Partnership for an Advanced Computing Environment (PACE) at the Georgia Institute of Technology (\verb|RRID:SCR_027619|).

\bibliographystyle{cas-model2-names}
\bibliography{cas-refs}

\end{document}